# Testing Origin-of-Life Theories with the Habitable Worlds Observatory

**Sukrit Ranjan**[1,2], **Martin Schlecker**[3,4], **Nicholas Wogan**[5], **Michael Wong**[6]

[1]*Lunar & Planetary Laboratory, University of Arizona, Tucson, AZ, USA*
[2]*Blue Marble Space Institute of Science, Seattle, WA, USA*
[3]*Steward Observatory, University of Arizona, Tucson, AZ, USA*
[4]*European Southern Observatory, Karl-Schwarzschild-Strasse 2, Garching by Munich, Germany*
[5]*Space Science Division, NASA Ames Research Center, Moffett Field, CA, USA*
[6]*Earth & Planets Laboratory, Carnegie Institution for Science, Washington, DC, USA*

The Habitable Worlds Observatory (HWO) aims to telescopically constrain the frequency and abundance of biospheres in the solar neighborhood. Origin-of-life theories vary in their predictions for the environmental requirements and the expected frequency of abiogenesis, meaning that constraints on the frequency and distribution of life on exoplanets from HWO can in principle directly test theories of abiogenesis. We categorize origin-of-life theories into three broad classes and discuss how HWO could potentially test them. Nondetection of biology on a large sample of habitable planets would provide prior-independent evidence in favor of the class of abiogenesis theory which holds that the origin of life is contingent, vanishingly unlikely event, whereas detection of event a single biosphere would falsify this class of theories. Correlations of candidate biospheres with planetary parameters such as UV irradiation, the presence of oceans, and the presence of continents can test specific origin-of-life theories. Simulated surveys with Bayesian analysis are required to quantify the ability of HWO to execute this science case. However, a clear theme from the limited such studies that have already been conducted is the need for large sample sizes ($\gtrsim 50$ planets characterized) to provide meaningful constraints on abiogenesis theories, favoring a larger design sample for HWO.

1. **Science Goal:** Test theories of the origin of life.

   A key goal of the Habitable Worlds Observatory (HWO) is to constrain the presence of life on exoplanets. But, how will this constraint advance our understanding of fundamental scientific questions? Here, we discuss how detection or nondetection of candidate biospheres with HWO can test theories of the origin of life, a fundamental scientific question since $\sim 500$ BCE (Catling 2013). Importantly, even nondetections of biology with HWO can constrain theories of abiogenesis (Rimmer 2023; Angerhausen et al. 2025). Simulated survey analysis in a Bayesian framework (e.g., Schlecker et al. 2025; Angerhausen et al. 2025) is required to quantitatively assess HWO's ability to test theories of abiogenesis.

2. **Science Objective:** Determine the frequency of candidate biosignatures and their correlation with environmental parameters (e.g., UV irradiation, presence of oceans) to test theories of abiogenesis.

   HWO may test theories of the origin of life by constraining the abundance and distribution of candidate biospheres (Chen & Kipping 2018; Ranjan et al. 2017; Kipping 2021; Rimmer et al. 2021; Rimmer 2023; Schlecker et al. 2025; Angerhausen et al. 2025):

   - Some theories of the origin of life suggest that abiogenesis is a rare event, and that there is no other biosphere within 10 pc (Rimmer 2023). For example, synthetic chemistry approaches to prebiotic chemistry require the interaction of aqueous reservoirs with distinct environmental conditions, e.g. by the interaction of streams. Requiring too many such interactions drives the probability of abiogenesis, and therefore of candidate biosphere occurrence, to a very low value (Rimmer 2023). Detecting even a single biosphere would falsify this family of abiogenesis theories, whereas ruling out biology on 20-50 planets with HWO would provide prior-independent evidence for an intrinsically unlikely abiogenesis (Angerhausen et al. 2025).

- Some theories of the origin of life predict that it is facile on habitable planets (e.g., Russell et al. 2014; Kipping 2020). For example, the alkaline vent theory (AVT) of abiogenesis suggests that the emergence of life is inevitable on any planet hosting a liquid-water ocean in contact with an alkaline hydrothermal vent and a $CO_2$-rich atmosphere, because life will necessarily emerge to resolve the thermodynamic disequilibrium generated by such a system (Russell et al. 2014; Sojo et al. 2016). Because oceans and $CO_2$ atmospheres may be abundant (Gaillard & Scaillet 2014; Ramirez & Levi 2018; Zeng et al. 2019), this would imply life is common. A high rate of occurrence of candidate biospheres in the HWO sample would favor facile abiogenesis theories as a class (of which AVT is one example), though discriminating between individual facile abiogenesis theories may be challenging.

- A wide range of theories of abiogenesis assert specific environmental requirements for the emergence of life. Best developed is the hypothesis that UV light is required for the emergence of life (Khare & Sagan 1971; Buccino et al. 2006; Rapf & Vaida 2016; Ranjan & Sasselov 2016; Ranjan et al. 2017; Rimmer et al. 2018; Rimmer et al. 2021; Green et al. 2021) Specifically, simulated surveys and Bayesian hypothesis testing reveal sample sizes of $\gtrsim 50$ planets can reveal correlations between past NUV irradiation and the presence of life, with the diagnostic power of the survey being strongest if the UV irradiation is in the range 200-400 ergs s$^{-1}$ cm$^{-2}$ (integrated 200-280 nm) (Schlecker et al. 2025).

For all of these possibilities, simulations need to be conducted to verify whether HWO has an adequate sample and prior knowledge to test these theories (e.g., Schlecker et al. 2025; Angerhausen et al. 2025).

## 3. Physical Parameters

- Determine presence of surface or atmospheric biosignatures and lack of false positive indicators This will be done as part of primary Living Worlds (biosignatures) science case)

- Determine presence of oceans and/or continents on planet. This is relevant to theories of the origin of life that require oceans and/or dry land, respectively (e.g., Russell et al. 2014; Deamer & Damer 2017).

## 4. Description of Observations

### 4.1. Observations

- High-precision UV-VIS spectra of temperate terrestrial planets in search of biosphere candidates.

- Spectral detection of $CO_2$.

- Phase-curve or spectroscopic observations in search of ocean glint (Robinson 2018).

- Phase-curve observations in search of continents (e.g., Meinke et al. 2022)

### 4.2. Complementary Observational Data

- UV and bolometric emission history of star (c.f. Richey-Yowell et al. 2023).

### 4.3. Suggestions for Future Work

- Need to conduct simulated surveys to determine the number and type of planets that must be characterized to obtain strong Bayesian evidence to confirm or reject these theories (e.g., Schlecker et al. 2025; Angerhausen et al. 2025).

## 5. Impact on Requirements

Table 1 provides the impacts of this science case on the requirements, following the HWO SSSCDD template.

**Acknowledgements.** SR gratefully thanks the University of Arizona for support via startup. MLW's research is funded by NASA through the NASA Hubble Fellowship Program Grant HST-HF2-51521.001-A awarded by the Space Telescope Science Institute, which is operated by the Association of Universities for Research in Astronomy, Inc., for NASA, under contract NAS5-26555.

Testing Origin-of-Life Theories    3Table 1: Impact on Requirements

| Capability or Requirement | Necessary | Desired | Justification | Comments |
|---|---|---|---|---|
| UV Observations | Yes | – | (A) Detect biosignature gas $O_3$. (B) Help constrain UV emission history of planet. | (A) already part of main Living Worlds science case. |
| Long wavelength observations ($> 1.5$ $\mu$m) | – | Yes | Detect biosignature gases (e.g., $CH_4$) and habitability indicators (e.g., $H_2O$) | Already part of main Living Worlds science case. |
| Timing of observations in different bands | Unknown | – | – | Phase-curve and/or rotation observations would be helpful to detect dry land/ocean presence. |
| High Spatial Resolution | No | No | Disk-integrated observations are assumed for this science case | |
| High Spectral Resolution | Unknown | – | May be useful to better model host star, infer UV emission history | |
| Large Field of View | No | No | N/A | |
| Rapid Response | No | No | N/A | |

Ramirez, R. M., & Levi, A. 2018, MNRAS, 477, 4627
Ranjan, S., & Sasselov, D. D. 2016, Astrobiology, 16, 68
Ranjan, S., Wordsworth, R., & Sasselov, D. D. 2017, The Astrophysical Journal, 843, 110. http://stacks.iop.org/0004-637X/843/i=2/a=110
Rapf, R. J., & Vaida, V. 2016, Phys. Chem. Chem. Phys., 18, 20067
Richey-Yowell, T., Shkolnik, E. L., Schneider, A. C., et al. 2023, ApJ, 951, 44
Rimmer, P. B. 2023, Origins of Life on Exoplanets (John Wiley & Sons, Ltd), 407–424
Rimmer, P. B., Ranjan, S., & Rugheimer, S. 2021, Elements, 17, 265
Rimmer, P. B., Thompson, S. J., Xu, J., et al. 2021, Astrobiology. http://doi.org/10.1089/ast.2020.2335
Rimmer, P. B., Xu, J., Thompson, S. J., et al. 2018, Science Advances, 4, eaar3302
Robinson, T. D. 2018, in Handbook of Exoplanets, ed. H. J. Deeg & J. A. Belmonte, 67
Russell, M. J., Barge, L. M., Bhartia, R., et al. 2014, Astrobiology, 14, 308
Schlecker, M., Apai, D., Affholder, A., et al. 2025
Sojo, V., Herschy, B., Whicher, A., et al. 2016, Astrobiology, 16, 181
Zeng, L., Jacobsen, S. B., Sasselov, D. D., et al. 2019, Proceedings of the National Academy of Science, 116, 9723